\documentclass[twocolumn,aps,pra,showpacs,superscriptaddress,
longbibliography,10pt]{revtex4-1}

\usepackage{bm}
\usepackage{amssymb}
\usepackage[usenames,dvipsnames]{color}
\usepackage{graphicx}
\usepackage{dcolumn}
\usepackage{hyperref}
\usepackage{natbib}
\usepackage[caption=false]{subfig}
\usepackage{float}
\usepackage{times}
\usepackage{booktabs}		
\usepackage{mathtools}
\usepackage{amsmath,array}
\usepackage{relsize}
\usepackage{enumerate}
\usepackage{color}

\usepackage{listings}
\usepackage{epstopdf}
\usepackage{braket}

\graphicspath{{figures/}}

\usepackage[utf8]{inputenc}

\begin{document}

\title{Finite-system effects on high-harmonic generation: from atoms to solids}

\author{Kenneth K. Hansen}
\affiliation{Department of Physics and Astronomy, Aarhus University, DK-8000, Denmark}
\author{Dieter Bauer}
\affiliation{Institute of Physics, University of Rostock, 18051 Rostock, Germany}
\author{Lars Bojer Madsen}
\affiliation{Department of Physics and Astronomy, Aarhus University, DK-8000, Denmark}

\begin{abstract}
Using time-dependent density functional theory, high-harmonic generation (HHG) is studied in one-dimensional structures of sizes from a single nucleus up to hundreds of nuclei.
The HHG cutoff is observed to extent linearly with the system size from the well known atomic HHG cutoff and is found to converge into the previously observed cutoffs for bulk solids only for large systems.
A change in the response from that of single atoms or small molecules is observed from system sizes of $N$ $\approx$ 6 nuclei and becomes independent of system size at $N \gtrsim 60$.
The system-size dependence of the observed HHG cutoffs is found to follow the limitations, set by the finite size solid, on the classical motion of electron-hole pairs. 
Because of the relation between recombination energy and electron-hole propagation length in the system, high-energy recombination events are not possible in small systems but become accessible for larger systems resulting in the change of the cutoff energies with system size. 
When varying the field intensity we observe that the cutoffs move linearly with the intensity even for small systems of $N\gtrsim 6$ that are far from the limit of a bulk solid.
\end{abstract}
\date{\today}
\maketitle

\section{Introduction}\label{intro}
High-harmonic generation (HHG) in gases has been studied extensively over the last three decades leading to new advances in production of short intense laser pulses \cite{atto_pulse_creation2} and new spectroscopic methods for study of atomic and molecular systems on a timescale of femto- and attoseconds\cite{attoKrausz}.
The emission of HHG from gases is well described by the three-step model consisting of (i) ionization of an atom, (ii) propagation of the freed electron in the external laser field and (iii) recombination of the electron returning to the atom and emission of the excess energy in terms of radiation \cite{Krause_1992,RecollCorkumPhysRevLett.71.1994}. 
This model predicts a cutoff in harmonic spectra at $\Omega_{\text{Max}} = 3.17 U_p + I_p$, where $I_p$ is the ionization potential of the atom, $U_p = F_0^ 2/4\omega_0^2$ is the ponderomotive energy here expressed as a function of the electric field strength, $F_0$, and the driving angular frequency of the laser pulse, $\omega_0$ (Atomic units are used throughout unless otherwise is indicated).
In molecular systems, molecular-specific effects have been predicted \cite{mol_interferens} and observed \cite{Kanai2005} in the HHG spectra, and processes such as molecular fragmentation from electronic excitation and ionization \cite{Molecule_destruction,molecular_chains} influence the HHG signal \cite{Molecule_destruction_pulse_length, mol_electronic_structure_from_hhg, HHG_from_organicmol}.
In atomic and molecular systems, additional cutoffs related to correlated two-electron dynamics have also been predicted \cite{NSDR,SPEAR_NSDR,HansenPhysRevA.96.013401} but these have very low signal strength.
As a recent example of the usage of the HHG process to gain insight in electron dynamics, HHG spectroscopy was used to resolve electron dynamics on the attosecond timescale \cite{Kraus790}.
 
Recently, there has been a growing interest in HHG from solid-state systems. 
This research has been partly focusing on using the solid-state systems to produce harmonics (see, e.g., Refs. \cite{GhimirePhysRevA.85.043836,Ndabashimiye2016}), and partly on using the emitted HHG radiation to probe ultrafast processes in solids (see, e.g., Refs. \cite{VampaPhysRevLett.115.193603,Schultze2013}).
Since HHG was first detected from a bulk solid-state system \cite{Ghimire2011} a few different solids have been used as media for the HHG process \cite{Luu2015,SchubertO.2014,Ndabashimiye2016} but also low dimensional systems \cite{SheetMaterialHHG} and systems with nanometer sized structures have been explored \cite{Nanoantennas}. 
Theoretically, the HHG process in bulk solids has quantum mechanically either been modeled using the semiconductor Bloch equations (see, e.g., Refs. \cite{Semiclassical_many_elec,HutterReviewLPOR:LPOR201700049,PhysRevLett.119.183902}), solving the time-dependent Schr\"{o}dinger equation for a periodic potential in an effective one-electron approximation \cite{gaarde_HHG,ishikawa_HHG,chinese_hhg_TDPI,Hawkins2015} or using time-dependent density functional theory (TDDFT) as a many-electron correlated approach in both reduced \cite{TDDFT_HHG} and full dimension \cite{Otobe0953-8984-21-6-064224,TancPhysRevLett.118.087403}, and recently including propagation effects \cite{PhysRevA.97.011401}.
From these studies it emerged that it is helpful to characterize solid-state harmonics as associated with either intra- or interband processes, even though it has been pointed out that there is no clear definition of intra- and interband HHG because of gauge dependence \cite{PhysRevB.96.035112}.
Intraband HHG is produced when electrons are driven in the bandstructure in bands that are not purely sinusodial. In this case the anharmonicity of the bands leads to the emission of higher harmonics \cite{Luu2015}. 
Interband HHG, on the other hand, is produced in a process that can be thought of as resembling the atomic three-step model with the following steps. (i) An electron tunnels to the conduction band producing a hole in the valence band, (ii) the electron and hole propagate in the solid and (iii) they recombine and emit the band gap energy at the crystal momentum at which the recollision occurs as HHG radiation\cite{PhysRevLett.113.073901}. 
For the interband harmonics, the maximum energy emitted in the HHG process is therefore defined by the path the electron takes in the band structure of the specific system under investigation.
In general, the HHG process in solids is more complicated than that in atoms because of the influence of the band structure on the HHG signal. In atomic HHG the electrons move in free space following the $E=k^2/2$ dispersion relation between the wavenumber, $k$, of the electron and its energy, whereas in solid-state HHG the electrons follow the dispersion relation given by the band structure.
For system sizes that are neither true bulk materials nor single atoms such as large molecules and nanometer sized structures the change from an atomic to a solid-state like HHG signal has not been examined and this is the purpose of the present work, where we also shed additional light on the HHG generation process in solids.

In this paper, we address the transition of the high-harmonic signal and generation mechanisms between atomic and solid-state systems. Using TDDFT \cite{UllrichBook,QSFQDBook}, we study the response of a range of system sizes from atomic systems to bulk solids observing the transition in the signal from atomic to solid-state HHG. In the intermediate region we address how the signal changes with system size which is of relevance for research in HHG from nanometer sized structures \cite{Nanoantennas,Cox2017}.
We find that the atomic HHG process is to some extend overtaken by the solid-state HHG process already at system sizes of $N \approx 6$ nuclei. From this system size, the HHG cutoff grows linearly in energy with system size from the atomic HHG cutoff towards the converged bulk solid-state HHG spectrum.
For intermediate system sizes between the atomic system and the bulk solid, we find that the finite size restricts the movement of the electron-hole pair limiting its propagation length to be smaller than the system size. 
This restriction of movement leads to a linear dependence of the HHG cutoff on the system size and also a linear dependence of the HHG cutoff on the pulse amplitude which earlier was found for bulk solids \cite{Ghimire2011} but now is shown also to be true for finite systems. 
These observations are explained using a classical three-step model for solid-state systems\cite{Semiclassical_many_elec,TDDFT_HHG,VampaTutorial0953-4075-50-8-083001} where the model reproduces the observed TDDFT cutoffs.

The paper is organized as follows. In Sec.~\ref{model} we introduce the TDDFT model. In Sec.~\ref{spectra} we present the HHG spectra obtained for different system sizes and discuss in detail the change from atomic HHG to solid-state HHG. In Sec.~\ref{sec:Classical_model} we analyze the HHG spectra using the extended classical electron-hole model \cite{TDDFT_HHG} to show that classical electron-hole movements in the band structure of the system results in the same maximum energy as that observed in the simulated spectra. Lastly, we summarize our results in Sec.~\ref{sum} and provide an outlook.

\section{(TD)DFT model}\label{model}
We consider many different system sizes but all consist of a linear chain of $N$ nuclei with a spacing of $a_0 = 7.0$ located at $x_i = [i -(N-1)/2]a_0$. The ionic potential has the form,
\begin{equation}
v_{\text{ion}}(x) = -\sum^{N-1}_{i=0} \frac{Z}{\sqrt{(x-x_i)^2 + \epsilon}},
\end{equation}
where $Z$ is the nuclear charge, and $\epsilon$ is the softening of the Coulomb potential here set to $\epsilon = 2.25$. We use $Z=4$ throughout, as discussed further below.

\begin{figure}
\centering
\includegraphics[trim={0 0 2cm 0},width=0.9\columnwidth]{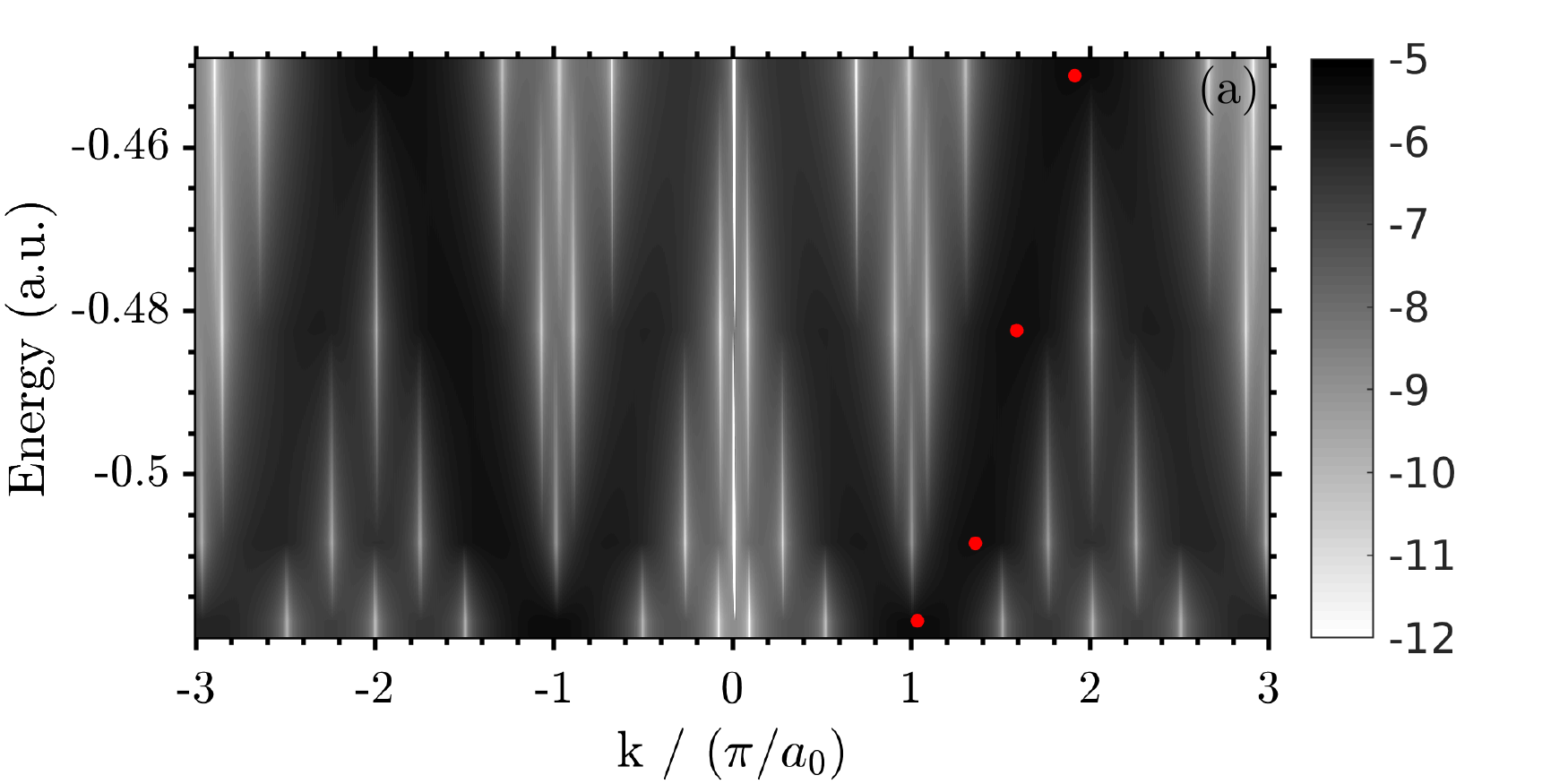}
\includegraphics[trim={0 0 2cm 0},width=0.9\columnwidth]{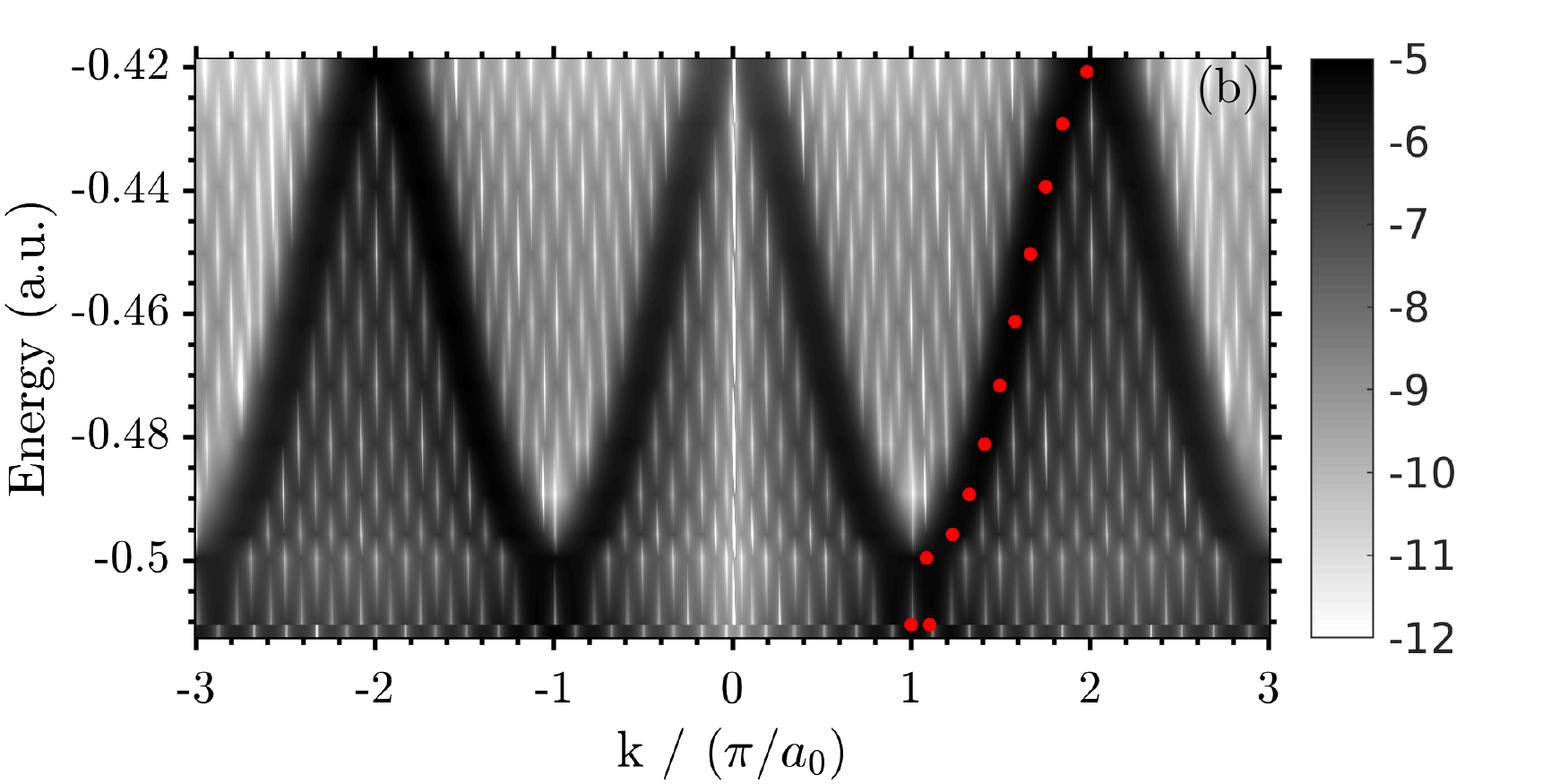}
\includegraphics[trim={0 0 2cm 0},width=0.9\columnwidth]{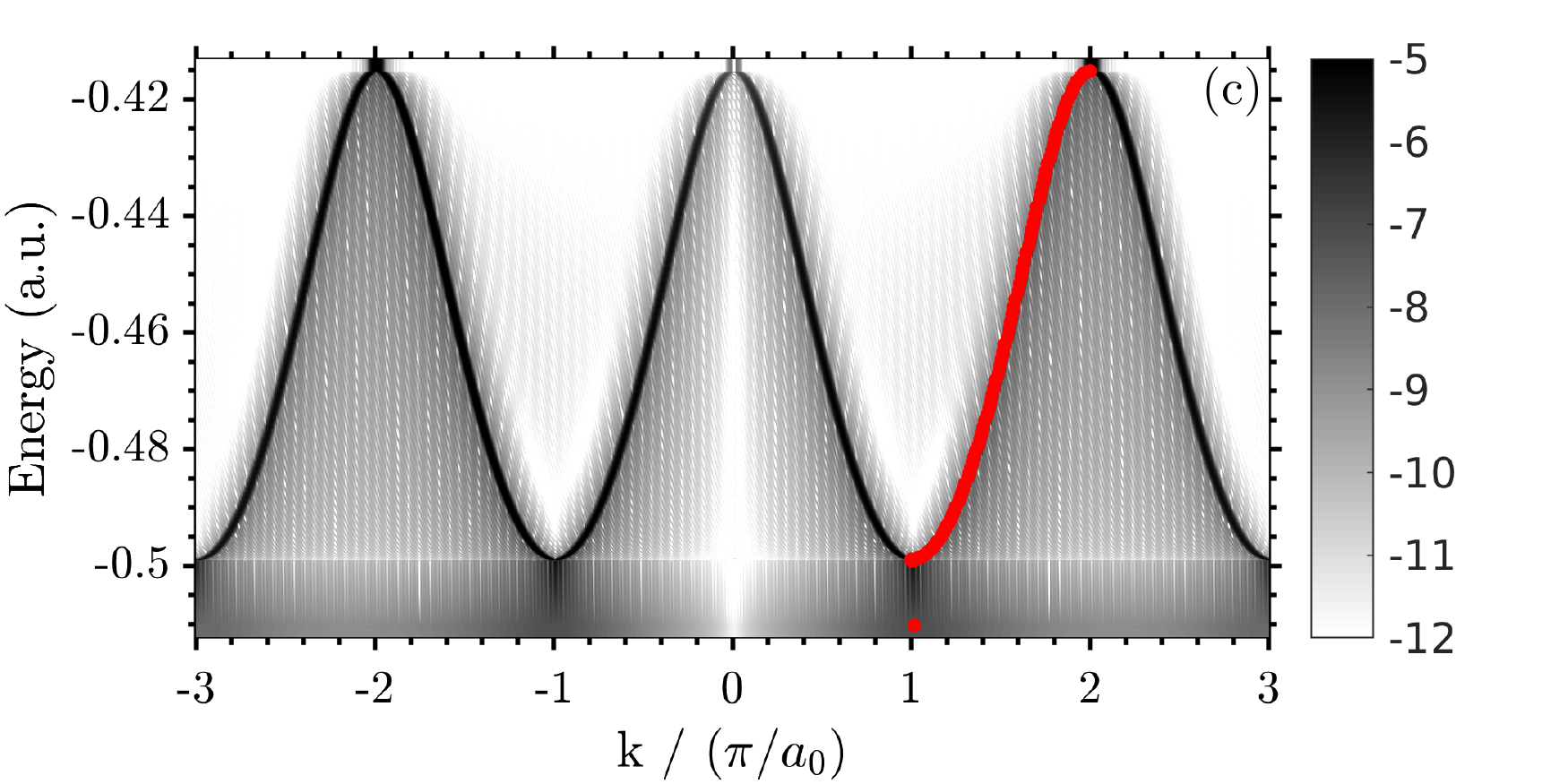}
\caption{Norm-squared of the Fourier transform of the eigenfunctions of the effective potential found from the DFT calculations (see text) for (a) $N=4$, (b) $N=12$ and (c) $N=100$ plotted at their respective energies on a logarithmic scale. Only the highest $N$ spin-up Kohn-Sham orbitals are shown. The maximum of the eigenfunctions in momentum space is indicated by (red) dots at the states energy.}
\label{fig:states}
\end{figure}

We find the electronic states for these systems using density functional theory (DFT), which enables us to simulate interacting electrons, at least on a mean-field level, and still be able to make calculations using hundreds of electrons. Our approach follows the method used in Ref. \cite{TDDFT_HHG}.
We use the ionic potential in the Kohn-Sham potential
\begin{equation}
v_{\text{KS}} [\{n_\sigma\}](x) = v_{\text{ion}}(x) + u[n](x) + v_{\text{xc}}[\{n_{\sigma}\}](x)
\label{eq:KSpot}
\end{equation}
which is then used for the calculation of the Kohn-Sham orbitals, $\varphi_{\sigma,i}(x)$, from the Kohn-Sham equation,
\begin{equation}
\epsilon_{\sigma,i} \varphi_{\sigma,i} (x) =  \left\{ -\frac{1}{2} \frac{\partial^2}{\partial x^2} + v_{KS}[\{ n_{\sigma}\}](x) \right\} \varphi_{\sigma,i} (x),
\label{eq:KS-eq}
\end{equation}
where also the Hatree potential
\begin{equation}
u[n](x) = \int \frac{n(x') dx'}{\sqrt{(x-x')^2 + \epsilon}},
\label{eq:hatreepot}
\end{equation}
and the exchange-correlation potential in a local spin-density approximation
\begin{equation}
v_{\text{xc}}[\{ n_{\sigma}\}](x) \simeq v_x [\{n_{\sigma}\}](x) = - \left[\frac{6}{\pi} n_\sigma (x) \right] ^{1/3},
\label{eq:ex-cor}
\end{equation}
has been used.
The spin densities and the total density are then found as
\begin{equation}
n_\sigma (x) = \sum ^{N_\sigma -1} _{i=0} |\varphi_{\sigma,i} (x) | ^2, \quad n(x) = \sum _{\sigma = \downarrow, \uparrow} n_\sigma (x).
\end{equation}
To retain spin neutral systems we set $N_{\uparrow,\downarrow} = Z  N/2$. We choose the same charge for all ions for all system sizes, with $Z = 4$ having a known result in the limit of a bulk solid \cite{TDDFT_HHG}. The number of orbitals, $\{\varphi_{\sigma,i}\}$, is chosen to make the system charge neutral, which makes the atomic limit of $N=1$ Beryllium-like. For small $N$, the system is molecular-like but with no real-world equivalent. For large $N$ the system converges to a solid-state bulk material with a band gap similar to real-world semiconductors.

We obtain groundstates for the different system sizes by propagating the Kohn-Sham orbitals in imaginary time using Eq. \eqref{eq:KS-eq} on a grid with a spacing of $\Delta x = 0.1$ where we orthonormalize the orbitals after each timestep.
We have determined the ground state for many systems of sizes of $N\in [1,100]$, focusing on systems with $N < 40$.
Using the Kohn-Sham potential in Eq. \eqref{eq:KSpot}, obtained for each system, we diagonalize the Hamiltonian in Eq. \eqref{eq:KS-eq} and obtain all states in the system.
The highest $N$ occupied states from the diagonalization are shown in Figs. \ref{fig:states} (a), (b) and (c) for $N=4$, $N=12$ and $N=100$, respectively. 
The states have been Fourier-transformed to momentum space and their norm-square are plotted at their respective energies on a logarithmic scale.
This produces a figure where the band structure of large systems will emerge as seen in Fig. \ref{fig:states} (c) but even for a small system such as $N=4$ in Fig. \ref{fig:states} (a) is it possible to see indications of a band-like structure. 
We will refer to these $N$ states in each system as the valence band. The maximum absolute value of the Fourier-transformed states are indicated with (red) dots. There are always as many states in a band as there are wells in the system but for small systems such as $N=4$ shown in Fig. \ref{fig:states} (a) there are not enough close-lying states to create a true band though there is already a signature of the band structure.
The plots of the Fourier-transformed states, determined from the diagonalization for unoccupied states, makes it possible to observe the different conduction bands. 

We propagate the orbitals obtained from the imaginary time propagation according to the time-dependent Kohn-Sham equation
\begin{align}
i\frac{\partial}{\partial t} \varphi_{\sigma,i} (x,t) = &\Big\{ -\frac{1}{2} \frac{\partial^2}{\partial x^2} -i A(t) \frac{\partial}{\partial x} \nonumber \\
&+ \tilde{v}_{\text{KS}} [\{n_\sigma\}](x,t) \Big\} \varphi_{\sigma,i} (x,t),
\label{eq:KS-eqTimedep}
\end{align}
where $A(t)$ is the vector potential of the laser pulse and where
\begin{align}
\tilde{v}_{\text{KS}} [\{n_\sigma\}](x,t) = v_{\text{ion}} (x) + u[n](x,t) + v_{\text{xc}} [\{n_{\sigma}\}] (x,t),
\label{eq:KSpotTimedep}
\end{align}
is the Kohn-Sham potential which is formally time-dependent because the time-dependent density enters the Hartree potential in Eq. \eqref{eq:hatreepot} and the exchange potential in Eq. \eqref{eq:ex-cor}. 
The time-dependent Kohn-Sham equation is solved using the Crank-Nicolson method \cite{QSFQDBook}.
Even for the intense pulses considered here, it was previously found that updating the Kohn-Sham potential in every timestep was not necessary as the total electron density hardly changes during the interaction with the laser pulse \cite{TDDFT_HHG} reflecting that most density remains in the initial ground state. We will therefore not be using Eq.~\eqref{eq:KSpotTimedep} for our calculations, but rather Eq.~\eqref{eq:KSpot} corresponding to the Kohn-Sham potential of the field-free initial state.

We assume a vector potential of the form
\begin{align}
A(t) = A_0 \sin^2\left(\frac{\omega_0 t}{2 n}\right) \sin(\omega_0 t),
\end{align}
with $\omega_0$ the driving frequency and $n$ the number of cycles. We use $n=6$ and $\omega_0 = 0.02305$ which is $\omega_0 = E_{\text{BG}} /10$, similar to what has been used experimentally \cite{Ghimire2011}, where $E_{\text{BG}}$ is the minimum band gap between the valence band and the conduction band for the $N = 100$-system. The value for $E_{BG}$ is $0.2305$ a.u., corresponding to $6.2$ eV.
For the TDDFT calculations we used $\Delta t = 0.1$ which is sufficient to converge the spectra. We calculate the expectation value of the current to determine the HHG spectra. The current is calculated as $j(t) = \sum_{\sigma,i} \int dx j_{\sigma,i} (x,t)$ with $j_{\sigma,i} (x,t) = -i [\varphi_{\sigma,i}^* (x,t) \partial_x \varphi_{\sigma,i} (x,t) - \varphi_{\sigma,i} (x,t) \partial_x \varphi_{\sigma,i}^* (x,t)]/2$ and the HHG spectrum is then given as the modulus square of the Fourier-transformed current $j(\Omega) = \text{abs}(\text{FFT}[j(t)])^2$ where a $\cos^8$ window function has been used on the current expectation value before Fourier transforming.
Identical spectra are obtained by considering the time-dependent dipole, $d(t) = \sum_{\sigma,i} \bra{\varphi_{\sigma,i}(x,t)}x\ket{\varphi_{\sigma,i}(x,t)}$, as $|j(\Omega)|^2 = \Omega^2 |d(\Omega)|^2$\cite{0953-4075-44-11-115601}.

\begin{figure}[ht!]
\centering
\includegraphics[width=1.0\columnwidth]{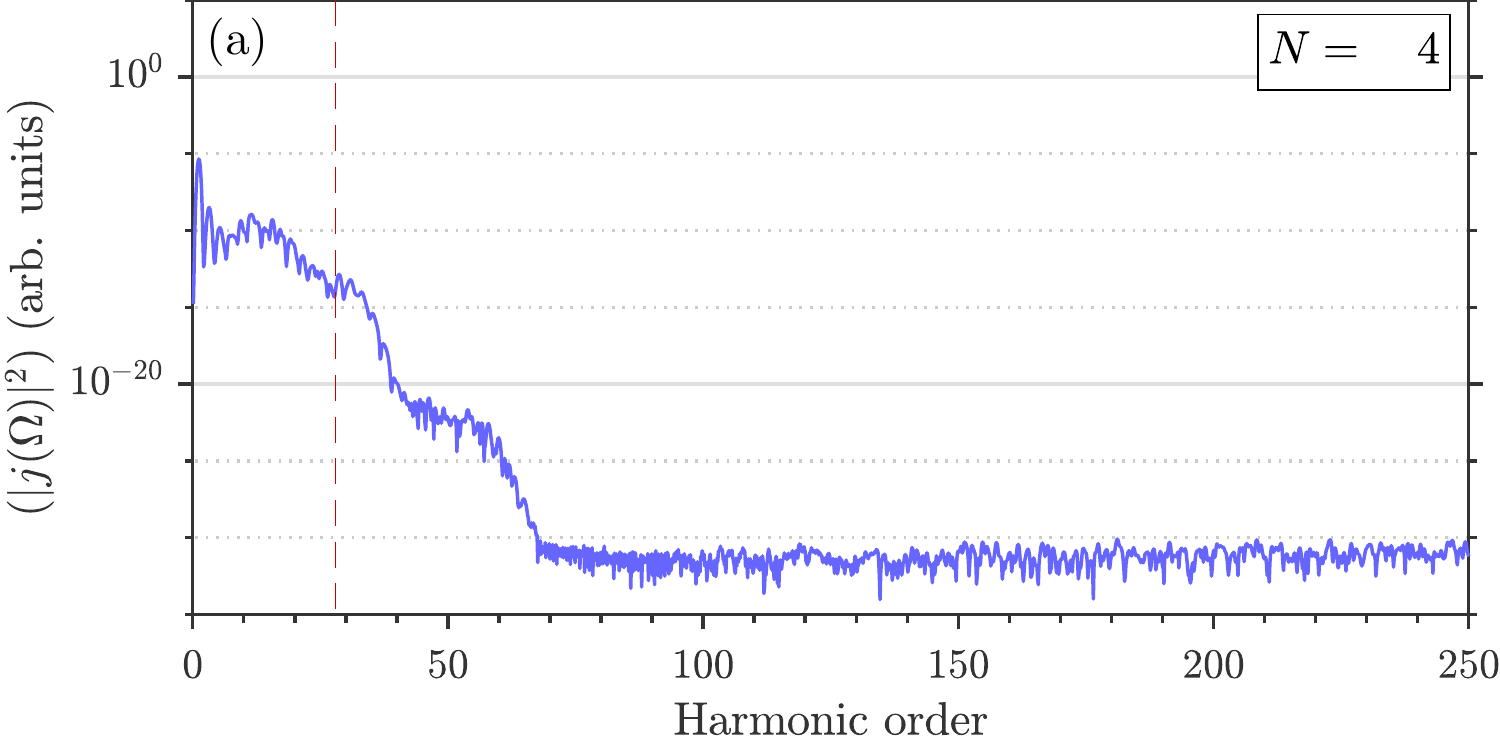}
\includegraphics[width=1.0\columnwidth]{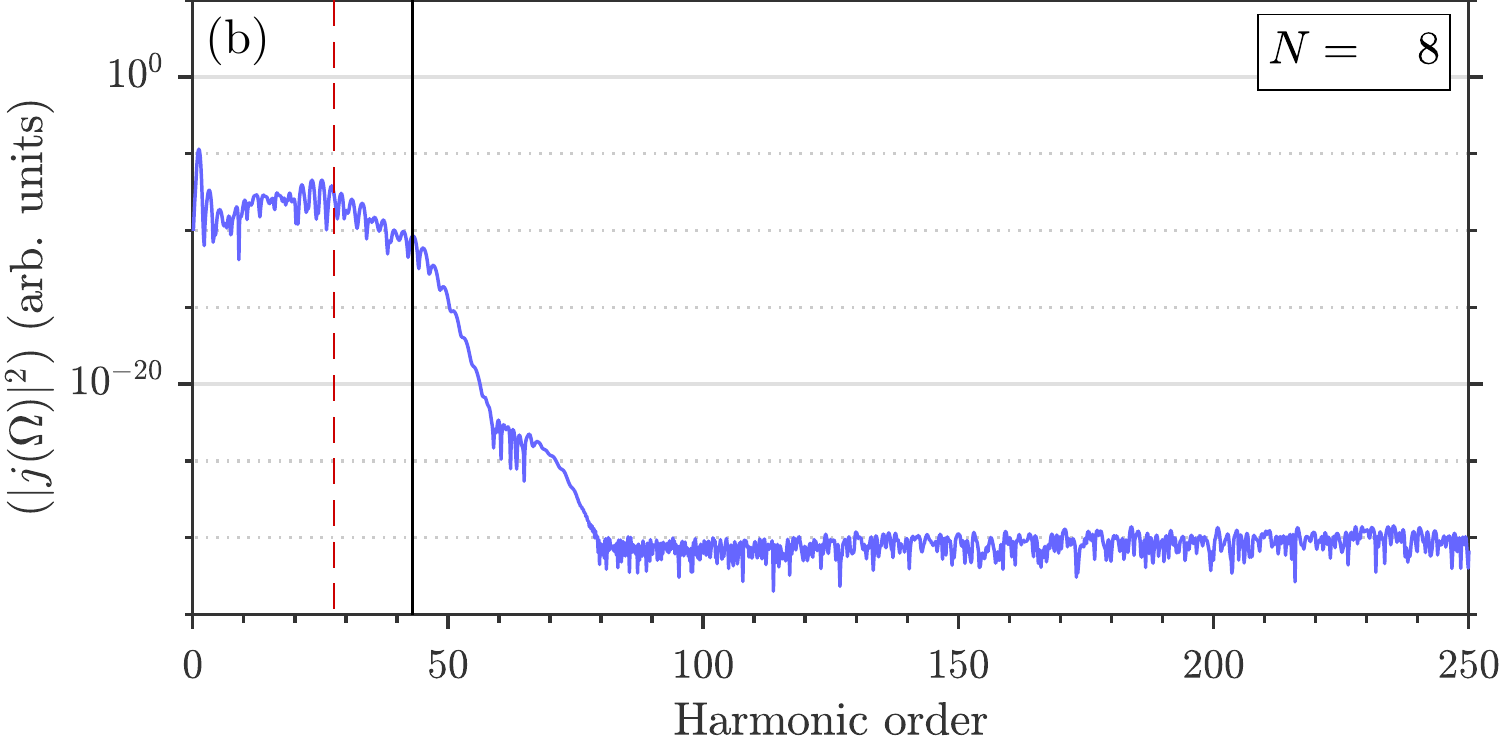}
\includegraphics[width=1.0\columnwidth]{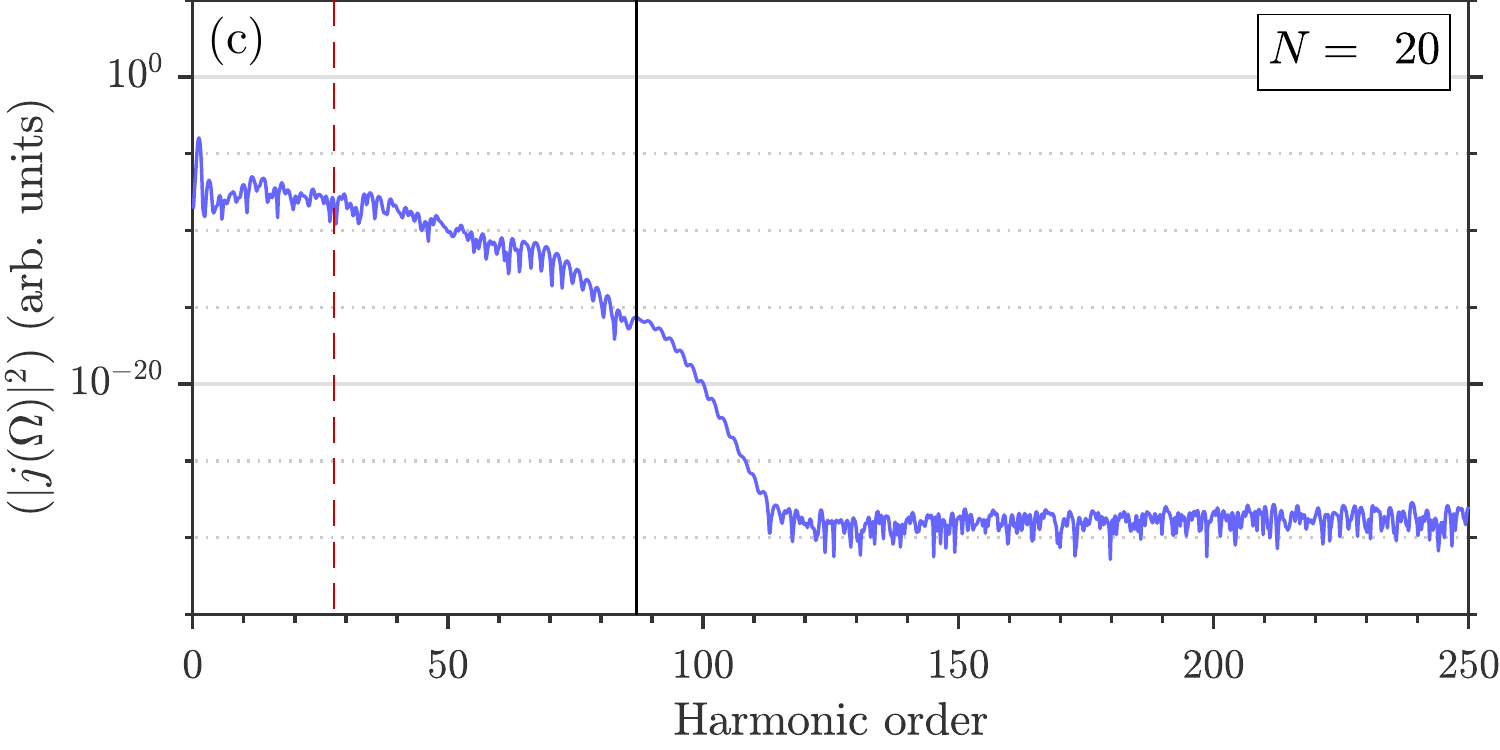}
\includegraphics[width=1.0\columnwidth]{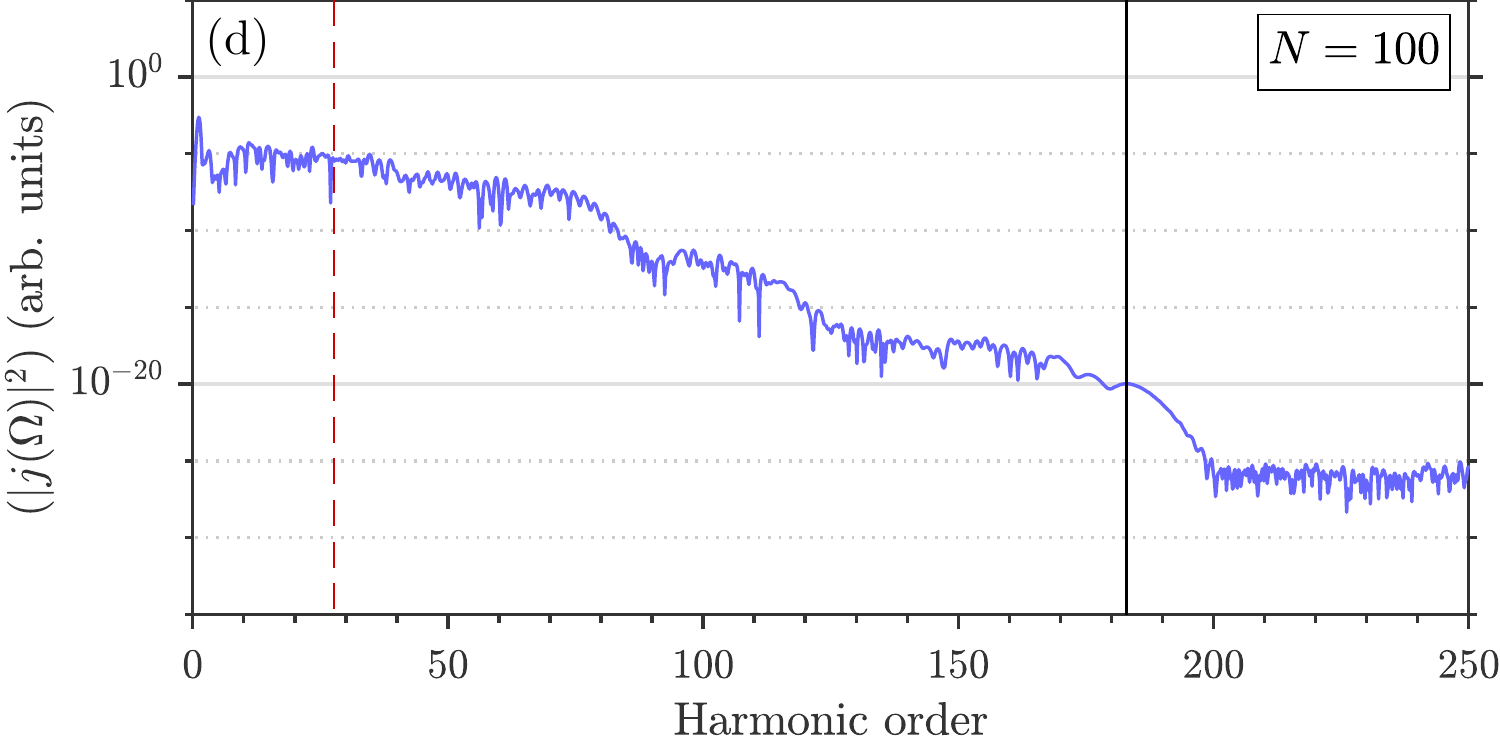}
\caption{HHG spectra for system sizes of $N=4,8,20,100$ nuclei in (a), (b), (c) and (d), respectively, using a $\sin^2$-pulse with $n=6$-cycles, $\omega_0 = 0.02305$ ($\lambda \approx 2 \mu$m) and $A_0 = 0.4$ ($I=1.6 \times 10^{13}$ W/$\text{cm}^2$). The vertical (red) dashed lines mark the atomic cutoff of $\Omega = I_p + 3.17 U_p$ and the vertical (black) full lines indicate the maximum observed cutoff for $N=8$, $N=20$ and $N=100$ in (b), (c) and (d).}
\label{fig:atomic}
\end{figure}

\section{From atomic to solid-state systems}\label{spectra}
To elucidate the transition from atomic-like to solid-state-like HHG, we first study the HHG spectrum from an atomic system i.e. $N=1$. For the atomic case (not shown) we observe the expected cutoff at $\Omega_{\text{max}} = I_p + 3.17 U_p$ where $I_p = 0.53$ for the case of $N=1$ in our system.
The atomic cutoff is present in the HHG spectrum up to a system size of $N=4$ which can be seen in Fig. \ref{fig:atomic} (a). For $N=4$ the atomic cutoff is located at a harmonic order of $\approx 30$ indicated by a vertical (red) dashed line. For systems of $N>1$, $I_p$ is set to the energy of the lowest state in the highest valence band. There is also a second plateau located at $\approx 50$ harmonics which originates from electrons in the lowest valence band in the system.
This is what is expected for our system mimicking a small molecule with atomic-like HHG spectra\cite{tdrnot4}.

For system sizes of $N\gtrsim 6$ the atomic cutoff law stops agreeing with what is observed for the cutoff in the TDDFT HHG spectra. For the system size of $N=8$, as presented in Fig. \ref{fig:atomic} (b), it is observed that the atomic HHG cutoff indicated by the dashed (red) vertical line is clearly not located at the cutoff for the system at the solid (black) vertical line.
This transition away from the atomic cutoff indicates that some other process than the atomic HHG process is active in the production of the HHG signal. When compared with a converged solid slab system as the case of $N=100$ in Fig. \ref{fig:atomic} (d) there is very little likeness between the $N=8$ and the $N=100$ spectra. 
The HHG spectrum for $N=100$ shown in Fig. \ref{fig:atomic} (d) holds several plateaus and cutoffs as is expected from a converged solid slab system \cite{TDDFT_HHG}. 
These features in the spectrum have previously been found to be associated with the band structure of the system and to result from an interband process where the recombination occurs from higher-lying conduction bands\cite{ishikawa_HHG}. There is in principle no upper limit to the energy that can be emitted from interband HHG. With a long enough pulse the electrons are able to move to an arbitratry conduction band, but as can be seen in Fig. \ref{fig:atomic} (d) the HHG signal reduces greatly for higher energies until it is not observable anymore. The highest energy cutoff in theoretical studies is therefore defined by the dynamic range of the calculation, which for our calculations is 25 orders of magnitude and thus much larger than in experiments.
Comparing the system with $N=8$ with states of $8 < N < 100$ it is found that the HHG cutoff increases approximately linearly with system size. In Fig. \ref{fig:atomic} (c) the intermediate case of $N=20$ is presented. There we can observe some of the characteristics of the $N=100$ system with a cutoff located around harmonic order 70 that can also be observed in Fig. \ref{fig:atomic} (d) but for $N=20$ in Fig. \ref{fig:atomic} (c) higher harmonics are suppressed by some yet unknown mechanism. We will address this point in Sec. \ref{sec:Classical_model}.

We expect that the transition away from atomic HHG that happens at system sizes around $N\gtrsim 6$ is connected with the increase in the phase space density in each band with increasing system size. 
The increase in the number of  states in the valence band enables a process similar to solid-state HHG to occur which gradually overshadows the signal from the atomic HHG process in the cutoff region. 
We observe a small amount of ionization for all system sizes so the atomic-like HHG process should still be present but the small amount of ionization would limit the atomic-like HHG process and suggests that only a small amount of atomic-like HHG is produced.
That the transition from atomic to solid-state HHG happens at such small system sizes is quite surprising as this size is comparable with small molecules or molecular chains. That such a transition has not been observed yet may be related to a destruction of molecules exposed to strong laser pulses and the atomic HHG cutoff being located at larger harmonics for the strong laser intensities, that were used for molecular systems \cite{HHG_molecules}.
Previous studies of atomic cluster systems with $N=3-9$ \cite{molecular_chains,atomic_clusters_TDDFT} also observed extensions of the HHG cutoff from the normal atomic HHG cutoff with system size but related it to the more energetic atomic HHG cutoff reported for extended molecular systems with a cutoff located at $\Omega_{\text{Max}} = I_p + 8U_p$\cite{ultra_HHG_mol}. This cutoff is lower than the cutoff we observe already at $N=8$ and therefore not causing the observed extension of the cutoff.

\begin{figure}
\includegraphics[width=0.8\columnwidth]{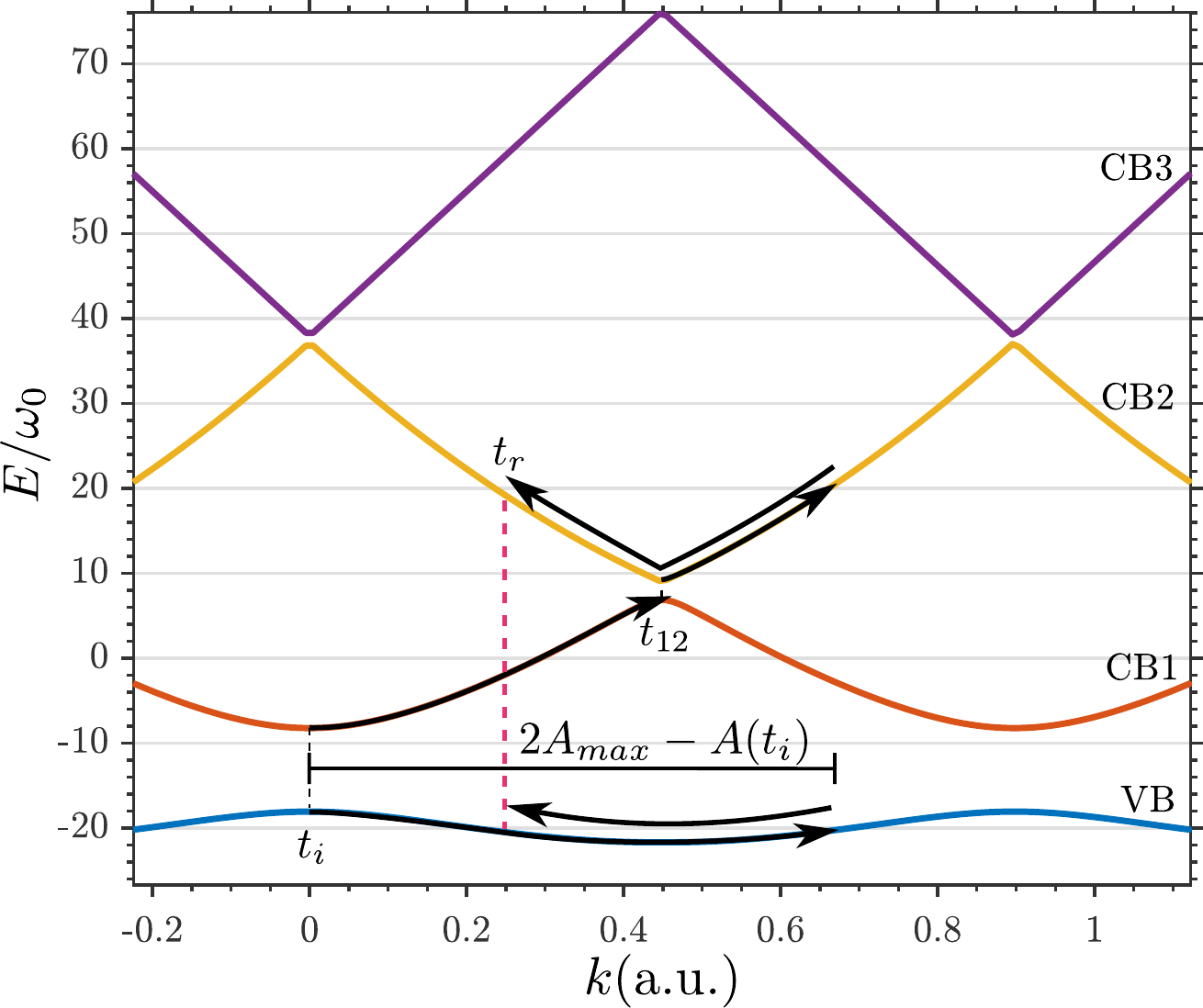}
\caption{The band structure of the $N=100$ system. The value of the minimum band gap is $0.2305$ a.u., corresponding to $6.2$ eV. The highest-lying valence band is shown and labeled as VB and the lowest three conduction bands are shown and denoted as CB1, CB2 and CB3. Indicated with solid (black) lines with arrows are the paths of an electron and a hole in the band structure from creation to recollision. The solid (black) line in the valence band is the hole moving adiabatically with the vector potential, with the amplitude of the movement being defined by the tunneling time, $t_i$. 
The vertical dashed (black) line at $t_i$, indicates the tunneling event of the electron at the minimum band gap between VB and CB1. The solid (black) lines with arrows in the conduction bands show the movement of the electron where at $t_{12}$ the electron jumps up to CB2, where it resides until recollision with the hole at $t_r$. At recollision a photon will be emitted with the energy of the band gap here indicated by the vertical dashed (pink) line.}
\label{fig:bandstruc}
\end{figure}

\begin{figure}
\centering
\includegraphics[width=1.0\columnwidth]{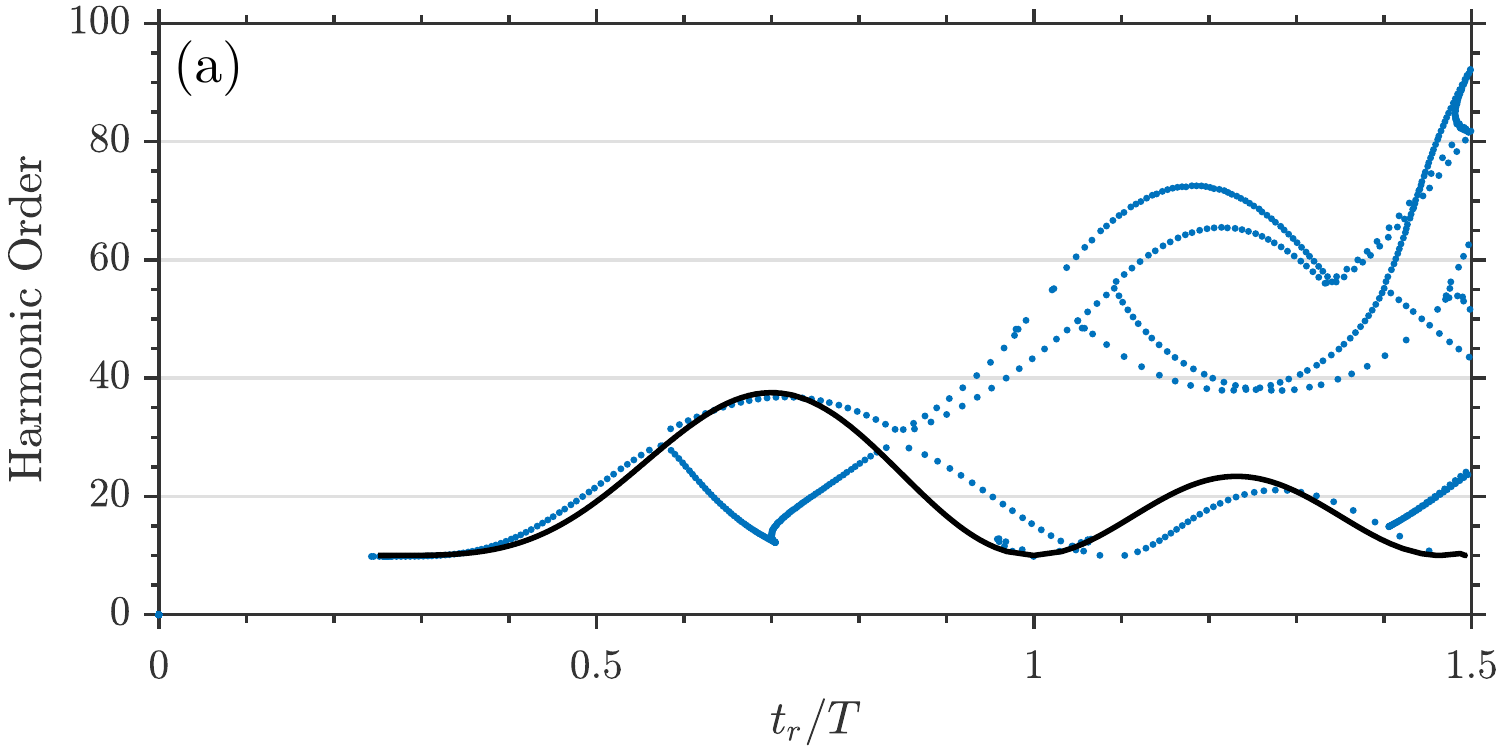}
\includegraphics[width=1.0\columnwidth]{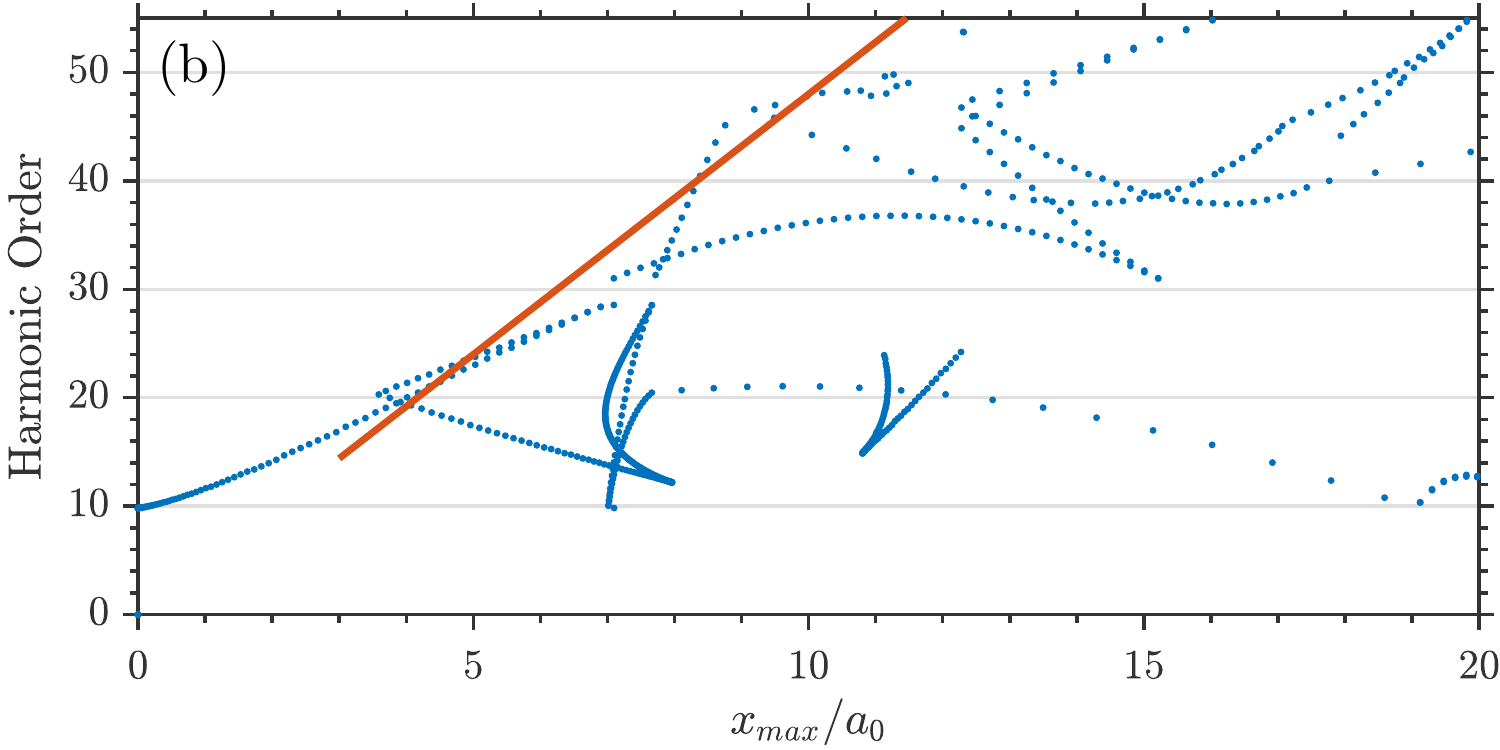}
\caption{Classical trajectories of holes and electrons in the band structure of $N=100$ for different tunneling times interacting with a continuous wave with $\omega_0 = 0.02305$ and $A_0 = 0.4$. (a) the band gap energy, expressed in terms of harmonic order, at the $k$ value at the instant of recollision as a function of recollision time over three laser half-cycles (dots). Also shown in (a) is the recollision energy in atomic HHG as a function of the recollision time found from the three-step model for atomic HHG (black solid curve). The curve has been scaled in amplitude to be similar to the solid state HHG recollision energy. (b) The band gap energy, expressed in terms of harmonic order, at the $k$ value where recollision occurs as a function of the propagation length of both the electron and hole for a given recollision event. The full (red) line in (b) is inserted to guide the eye. The line indicates that from a certain harmonic order, say 15, there is approximately a linear relation between the minimal distance between the electron-hole pair that results in a given harmonic order and this harmonic order (see text).}
\label{fig:classical}
\end{figure}

\section{Model Analysis of Finite system effects on HHG}\label{sec:Classical_model}
When comparing the cases $N=8$, $N=20$ and $N=100$ in Figs. \ref{fig:atomic} (b), (c) and (d) it is observed that the HHG cutoff grows almost linearly with system size, which is also a trend supported by our findings for all other intermediate system sizes studied.
A linear dependence of the HHG cutoff was observed for the external field amplitude in the first experimental realization of solid-state HHG in bulk solids \cite{Ghimire2011} and a semiclassical approach was able to describe the process leading to this linear dependence \cite{Semiclassical_many_elec}.
Using an extended version \cite{TDDFT_HHG} of the semiclassical model\cite{Semiclassical_many_elec}, we will analyze the observed dependence of the HHG cutoffs on the system size in our TDDFT calculations.
The model will be explained in the following.

We assume the creation of an electron-hole pair at time $t_i$ in the valence and first conduction band, respectively. Note that we use the words valence and conduction band, also in the case of relatively few nuclei. This usage is inspired by Fig. \ref{fig:states}, which shows traces of what one could refer to as a discretized band structure already for $N=4$. To be able to evaluate the equations of motion [Eq. \eqref{eq:ESCHE} below], we will fit a continuous curve to these discrete valence and conduction bands. No trajectories are omitted in accounting for finite system effects. Rather, as we shall see, trajectories with maximum extension larger than the system size do not contribute as much to the HHG signal in the cutoff region as trajectories that produce solid-state-like harmonics within the finite system. 
The momenta of the electron and hole will then be given by the relation $k(t) = k_0 + A(t)$\footnote{The momentum of the hole is in principle $k(t) = -k_0 - A(t)$ but the energy of the hole also changes as the inverse of an electron in the valence band and does not follow the valence band as assumed. These two faulty assumptions cancel each other out so that the ESCHE model still holds true\cite{Kittel_solid-state}}, where $k(t)$ is the time-dependent momentum, $k_0$ is the initial momentum at the minimum band gap between valence band and conduction band which for our system is zero and $A(t)$ is the vector potential of the external field. Using the time-dependent momentum, $k(t)$, we find the velocity of the hole and electron from the slope of the valence and first conduction band, respectively (see Fig. \ref{fig:bandstruc} there abbreviated as "VB" and "CB1").
The position of the electron or hole is then given as:
\begin{align}
x(t) - x_0 = \int^t_{t_i} \nabla_k E [k(\tau)] d \tau,
\label{eq:ESCHE}
\end{align}
where $x_0$ is the initial position of the hole and electron which both are located at $x_i = 0$ at $t_i$ and $E [k(\tau)]$ is either the valence or conduction band depending on whether the hole or electron is considered. Similar to the three-step model of HHG in gases we assume the electron and hole to propagate independently until they recollide at some time $t_r$.
The energy released then equals the band gap between the valence and conduction band at $k(t_r)$ in the band structure. 
Only including two bands in the model works for small field strengths but when the vector potentials moves the electrons in the conduction band to the Brillouin zone boundary, located at $\pi/a_0 = 0.45$, we need to enable the electrons to jump to higher conduction bands [Fig. \ref{fig:bandstruc}]. The inclusion of several conduction bands was previously also used in Ref. \cite{TDDFT_HHG} where it was called the extended semiclassical hole-electron model (ESCHE).
The jumping between conduction bands is illustrated in Fig. \ref{fig:bandstruc} where at $t_{12}$ the electron reaches the minimum band gap between the first and second conduction band and then moves from CB1 to CB2. At $t_{12}$ the electron moves up into the CB2 without disturbing the hole, which is still located in the valence band and continues to move according to it and Eq.~\eqref{eq:ESCHE}.
After the jump the electron moves according to the dispersion relation of CB2. We allow jumps between subsequent conduction bands at the $k$'s of all minimum band gaps. The emitted energy at recollision is then equal to the band gap between the conduction band the electron moves in and the valence band at $k(t_r)$ indicated by the vertical dashed (pink) line in Fig. \ref{fig:bandstruc}. In Fig. \ref{fig:bandstruc} the electron and hole move first to the right with the vector potential and then move to the left after the vector potential reaches its maximum amplitude because, as in the atomic case, there can be no recollision events before the electron changes direction.

\begin{figure}
\includegraphics[width=1.0\columnwidth]{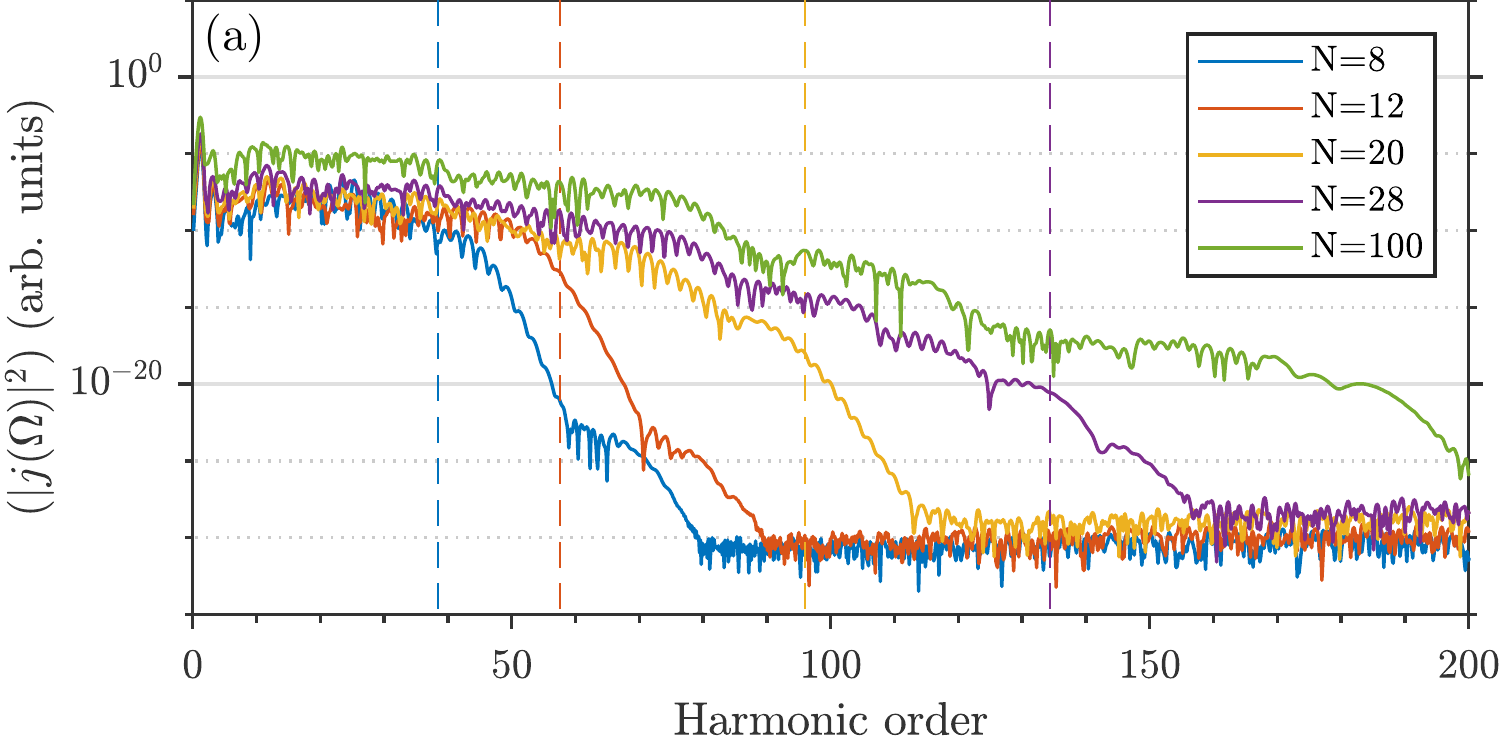}
\centering
\includegraphics[width=1.0\columnwidth]{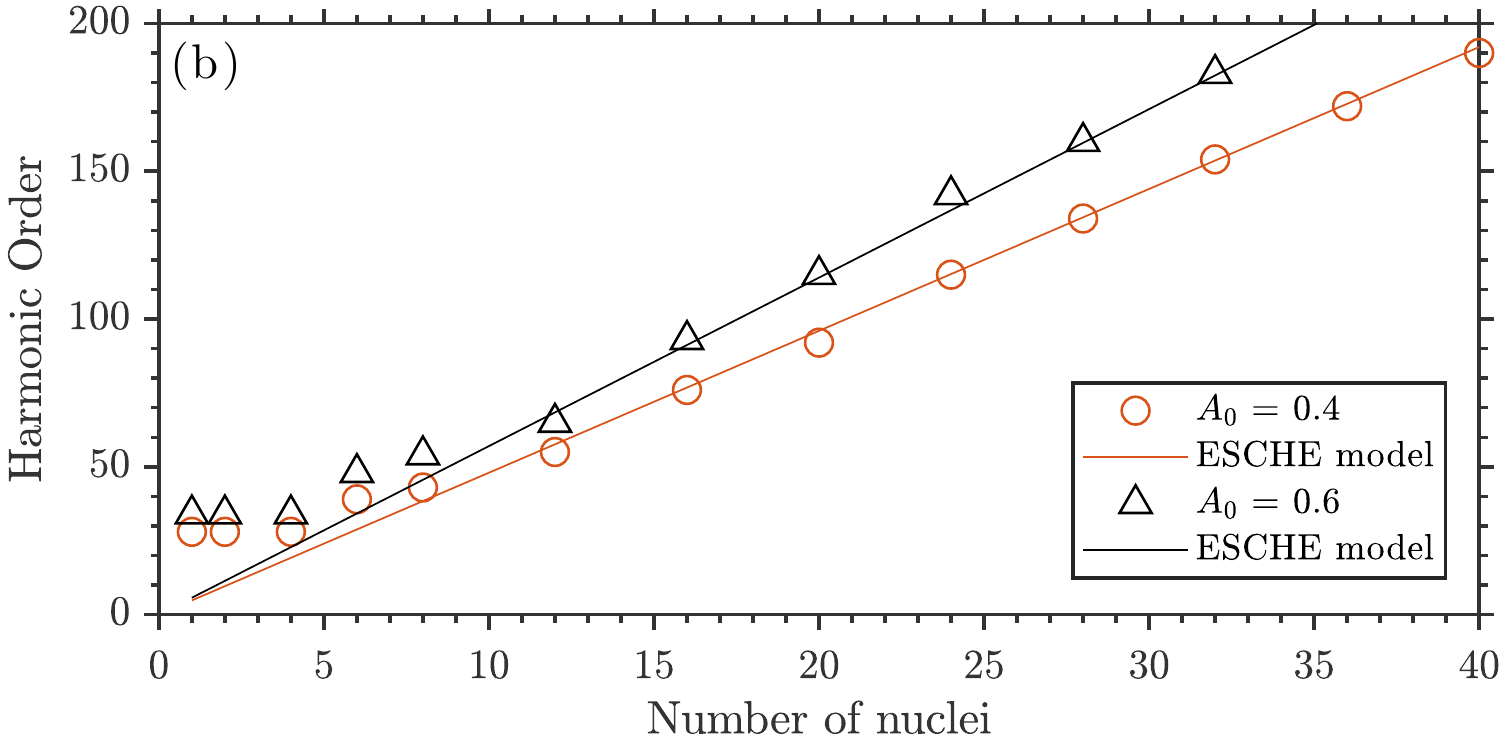}
\caption{(a) HHG spectra for system sizes of $N=8,12,20,28,100$ using the same pulse parameters as in Fig. \ref{fig:atomic}. The HHG spectra are plotted with growing system size from left to right. The vertical dashed lines are the HHG cutoffs predicted by the ESCHE model which also grow with system size from left to right. The spectrum and vertical dashed line of the same color fit together. (b) The observed cutoffs as a function of system size for vector potential amplitudes of $A_0 = 0.4$ and $A_0 = 0.6$ are plotted with the linear fit of the HHG cutoffs found from the ESCHE model for each external field amplitude.}
\label{fig:combined_HHG}
\end{figure}

Using the state energies of the $N=100$ [Fig. \ref{fig:states} (c)] system we fit a continuous curve to the state energies in each band and create a band structure and then calculate the recollision times for holes and electrons for different tunneling times $t_i$. 
The band gap energy, expressed in terms of harmonic order, at recollision time as a function of the recollision time is shown in Fig. \ref{fig:classical} (a). 
In Fig. \ref{fig:classical} (a) each dot represents a recollision event and if infinitely many tunneling times were probed, the dots would form continuous lines. 
Each time, $t_i$, can lead to several recollision times because of the jumping between conduction bands leading to a more complicated relation between the recollision time and the energies that can be emitted. 
As a comparison we have included the atomic equivalent in Fig. \ref{fig:classical} shown as the solid black curve, which is obtained using the three-step model for HHG in atoms. 
The atomic recollision energies have been scaled in amplitude to be comparable with the ESCHE model calculations.

To study finite size systems we plot in Fig. \ref{fig:classical} (b) the recollision band gap energies, again expressed in terms of harmonic order, as a function of the combined maximum distance the hole and electron move i.e. $x_{max} = \textrm{abs}(x_{\textrm{Hole}})_{max} + \textrm{abs}(x_{\textrm{Electron}})_{max}$.
After creation, the electron and hole will move away from each other until they change direction enabling recollision.
Different $t_i$ can lead to recollision events with similar emitted energies but from different paths as seen in Fig. \ref{fig:classical} (b) where for a certain harmonic there can be many recollision events. For this study what is of interest is the paths with the smallest $x_{max}$ for a certain harmonic as this will be the smallest size a system can have to be able to emit this harmonic.
In Fig. \ref{fig:classical} (b) it is clear that higher harmonics have a larger minimum propagation distance before recollision therefore needing a larger system to contain such a motion. In Fig. \ref{fig:classical} (b) a (red) line has been plotted approximately over the left most points along the minimal distance required for the respective harmonic order in the plot which shows why the HHG cutoff grows approximately linearly with system size.

That the ESCHE calculations were made from the band structure of the $N=100$ system could have an effect on the classical results since it was seen in Fig. \ref{fig:states} that the band structures change significantly with the number of nuclei. 
When finding the band structure for the classical calculations one would, however, have to fit a line to the band structure of the finite systems and this line would be almost identical to the one determined for the $N=100$ system. There should therefore not be a significant difference in the classical calculations if one used the band structure for the respective system sizes.

In Fig. \ref{fig:combined_HHG} (a) the HHG spectrum for several system sizes are plotted together with the predicted cutoff from the ESCHE model. We note that a given $N$ corresponds to a spatial extend of the system with the size $N a_0$, with $a_0$ the spacing between nuclei. The cutoff in the spectra can be estimated in the ESCHE model requiring $x_{max} \lesssim N a_0$ [Fig. \ref{fig:classical} (b)]. The vertical dashed lines indicate the classical maximum recombination energy of the electron-hole pair for a certain system size and the colors of the vertical line fits with the colors of the spectra. 
It is observed that the linear fit of the classical calculation follows the HHG cutoff very well for increasing system size. 
It is important to make the distinction between the usual cutoffs observed in HHG spectra for atomic and solid state bulk materials and the cutoffs for finite systems.
The cutoff for atomic and solid-state bulk system HHG originates from the limitation of the classical path's maximum recombination energy creating a sharp drop. In contrast in the finite solid-state system the signal is instead dampened beyond the classically allowed energies creating less pronounced cutoffs.

The ESCHE calculations have also been done for other field amplitudes and the same agreement between quantum and classical calculations was found as can be seen for $A_0 = 0.4$ and $A_0 = 0.6$ in Fig. \ref{fig:combined_HHG} (b) where the observed cutoffs are plotted with the classically predicted linear dependence of the HHG cutoff.
These calculations make it clear that the HHG cutoff observed for system sizes of $N\gtrsim6$ is related to the solid-state slab HHG process but the system sizes restrict the maximum recombination energy of the electron-hole pair.
The fact that electrons or holes reaching the edge of the system do not contribute to the HHG signal is not trivial. They might still recollide leading to a signal in the HHG spectrum but no such structures are observed. 
When comparing the $N=100$ spectrum in Fig. \ref{fig:combined_HHG} (a) with the smaller system sizes it is observed that all structures in the spectra are either also present in the $N=100$ spectrum or they are created from the damping of the signal after the ESCHE model-predicted cutoff. 
As an example of the above, we notice from Fig. \ref{fig:combined_HHG} (a) that the cutoff in the $N=100$ spectrum around 75 harmonics can also be observed in the $N=20$ and $N=28$ systems.
There is therefore no indication that electrons or holes reaching the edge of the system contribute to the HHG signal. This is different from edge states that have been found to contribute significantly to the HHG signal when present\cite{topological_edge_states}.
In real bulk solids there will be domains of crystal structure with boundaries where the symmetry of the system is broken. These domains could act as separate nanometer sized systems where the HHG signal is reduced because they are not large enough to contain the solid-state HHG process. For the amount of harmonics that is experimentally detected at this time, the domain size does not appear to have been a restriction.

\begin{figure}
\centering
\includegraphics[width=1.0\columnwidth]{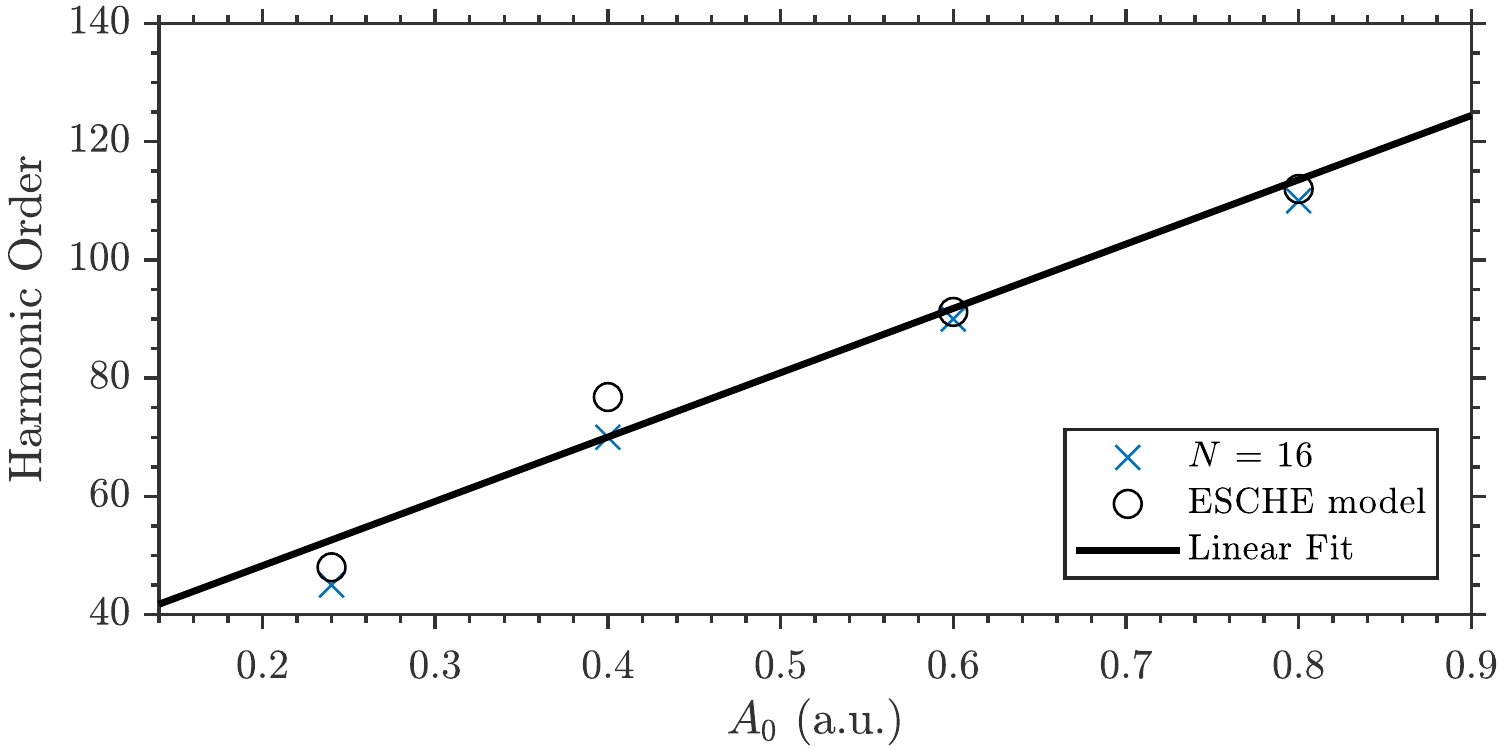}
\caption{HHG cutoffs for a system size of $N=16$ as a function of vector potential amplitude, $A_0$, using the pulse parameters of Fig. \ref{fig:atomic} are shown as crosses. Using the ESCHE model, a dependence on $A_0$ for the $N=16$ system was determined as shown by circles. A linear fit of the ESCHE model points is included.}
\label{fig:cutoff_int}
\end{figure}

We also tested the dependence of the HHG cutoff on the magnitude of the vector potential amplitude, $A_0$. The ESCHE model depends strongly on $A_0$ and the linear fit of the minimum propagation distance (solid (red) line in Fig. \ref{fig:classical} (b)) therefore changes with $A_0$. The observed cutoffs for the system size of $N=16$ as a function of $A_0$ is plotted in Fig. \ref{fig:cutoff_int}. 
It is observed that finite solid-state systems depend linearly on $A_0$.
Also plotted in Fig. \ref{fig:cutoff_int} is a linear regression of the cutoff found from the ESCHE model for $N=16$.
From the ESCHE model it is concluded that the higher field amplitudes enables electrons to move higher in the band structure without moving further in real space. 
This depends linearly on the field, which is why we observe a linear dependence on $A_0$ in contrast to the case of a bulk solid where cutoffs move linearly with field amplitude because of the shape of the band structure \cite{Semiclassical_many_elec}.

\section{Summary}\label{sum}
Using a self-consistent TDDFT model, we have addressed the transition from atomic-like systems to solid-state bulk materials with respect to the emission of HHG and in particular the cutoff. We observe a transition away from the well known atomic HHG cutoff already at system sizes of $N\approx 6$ nuclei. For $N\gtrsim 6$ we observe a continuous change of the cutoff with growing system size.
The change from atomic to solid-state HHG at such small system sizes is proposed to stem from the changes in the state density in momentum space. 
For $N \gtrsim 6$, the increase in the number of states in the valence band enables a harmonic generation process similar to the one in bulk materials, and with increasing $N$ this gradually overshadows the signal from the atomic HHG process. 
For $N > 6$ the solid-state HHG process, which has less of an energy gap for tunneling, will dominate the cutoff region of the HHG signal.
This would indicate that small molecules and nanometer sized systems should be able to emit solid state HHG opening up the possibility to use such systems as antennas for the solid state HHG process\cite{SheetMaterialHHG,Nanoantennas}. Conversely the solid state HHG process could be used to probe properties of such small systems on a femtosecond timescale.

The transition from atomic to solid-state HHG could hint at a possible different solid-state HHG process for small systems but further studies using an extended semiclassical hole-electron model revealed that the HHG signal in fact originates from the normal bulk solid-state HHG process but with the limitations of the real-space movement of the hole and electron in the solid. This limits the emission of high-frequency HHG from small systems. 
A linear dependence of the HHG cutoff on the vector potential amplitude was found.
This dependence of the solid state HHG process on the system size shows how crystal domain sizes could be a limiting factor for emission of very high harmonics from the solid state systems. The system size dependence of the solid state HHG process also opens up for the study of the spatial shape of an object with the HHG signal.

\section*{Acknowledgments}
K.K.H. and L.B.M. acknowledge support from the Villum-Kann Rasmussen (VKR) center of excellence QUSCOPE - Quantum Scale Optical Processes.


%

\end{document}